\documentclass{aa}
\usepackage{graphicx}
\begin{document}
\title{RR Lyrae stars in the inner Large Magellanic Cloud: Halo-like location with a disk-like distribution}
\author{Annapurni Subramaniam\inst{1}}
\offprints{Annapurni Subramaniam}
\institute{Indian Institute of Astrophysics, Koramangala II Block, Bangalore - 34\\
            \email{purni@iiap.res.in}}
\date{Received / accepted}
\titlerunning{RR Lyrae stars in the inner LMC}
\authorrunning{Subramaniam}
\abstract
{}
{
The distribution of RR Lyrae stars (RRLS) in the inner Large Magellanic Cloud (LMC),
and the structure of the halo of the LMC delineated by these stars are studied here.
}
{RRLS identified by the OGLE II survey 
are used to estimate their number density distribution in the bar region of the LMC. 
To find their location, I estimated the scale-height of their distribution
in the LMC using extinction corrected average magnitudes of ab type stars. }
{The density is found to vary differently along and across the bar of the LMC,
and the difference is found to be statistically significant. The density distribution
is found to be elongated like the LMC bar and the position angle 
(PA) of the elongation is estimated to be
112.$^o$5 $\pm$ 15.$^o$3. This value of PA is found to be same as the 
PA$_{maj}$ of the bar, within the errors,
estimated using red clump stars and giants. The ellipticity of their
density distribution is estimated to be $\sim$ 0.5, very similar to the
ellipticity of the bar, estimated from giants. 
The above results show that majority of the population of RRLS in the
central region of the LMC are found to have the signature of the bar. 
This result could mean that most of these stars
are located in the disk, considering the bar as a disk feature.
On the other hand, their scale-height was found to be 3.0$\pm 0.9$ kpc. This indicates that RRLS are located
in the halo and not in the disk. }
{Thus these stars in the inner LMC have halo-like location and
a disk-like density distribution. I discuss some possible formation scenarios
for this puzzling combination.}
\maketitle
\section{Introduction}
The Large  Magellanic Cloud (LMC) is known to be a disk galaxy with or without a halo.
There have been many efforts to find evidence for the presence of a halo in the LMC.
Such evidence is looked for in the old stellar population like the globular clusters (GCs) and
the field RR Lyrae stars (RRLS). The oldest GCs appear to lie in a flat rotating disk whose
velocity dispersion is 24 km s$^{-1}$ (Kinman et al. 1991 and Schommer et al. 1992).
van den Bergh (2004) re-estimated the velocity dispersion of the GCs and concluded that they
could still have formed in the halo. Since the number of GCs in the LMC is small 
(13 clusters), it is difficult to infer the signature of the halo from this
sample. The other tracer is the RRLS, which are almost as old as the oldest GCs.
Recent surveys of the LMC like the OGLE and MACHO have identified a
large number of RRLS. Among the follow up studies, Minniti et al. (2003) found a kinematic 
evidence for the
LMC halo, by estimating the velocity dispersion in the population of RRLS. Alves (2004) estimated
the mass of the LMC based on their surface density and remarked that their
exponential scale length is very similar to that of the young LMC disk.
Freeman (1999) remarked that the similarity of their scale length
with the scale length of the young stellar disk suggested that 
the RRLS in the LMC were disk objects, like the old clusters, 
and supported the view that the LMC may not have a metal-poor halo.

In this study, I used the RRLS catalogue of Soszynski et al. (2003) in the LMC, consisting of
7400 stars (after correcting for multiple entries), to obtain their number density distribution. 
The aim of the present study
is to identify any feature in the distribution, that could be described as the signature
of a halo or a disk distribution.
Indications for a distribution which is not circularly symmetric in the inner regions can be 
seen in figure 12 of
Alcock et al. (1996), where a large fluctuation in the RRLS density is found in the
inner 3 Kpc. 
The above fluctuations could arise due to averaging in position angles (PAs), when there is
a non-circular distribution. The density map of RRLS
shown in figure 1 of Soszynski et al. (2003) indicated that the stars have a preferred distribution
along the major axis of the bar, with a steep gradient across the bar. This is a
strong indication that these stars may be located in an elongated bar like distribution.
The RRLS as detected in the OGLE survey are projected on the bar, 
since the OGLE survey essentially traces the bar. If they belonged to the
halo, then their distribution in the inner LMC should show a symmetric distribution.
On the other hand,
if their distribution showed differences along and across the bar and also any other
property of the bar as found in the younger population, then the 
RRLS population in the inner LMC may belong to the bar or disk, since the bar is considered as a disk
feature. This point is further explored in this study by comparing the 
distribution of RRLS with the distribution of red clump (RC) stars and red giant stars, where 
RC stars and giants belong to a younger population and also trace the feature of the LMC bar.

If the distribution of RRLS mimics that of the bar/disk, then the obvious conclusion will be that
the stars are located in the disk of the LMC and not in the halo. This conclusion needs to be verified
by finding the actual locations RRLS. This can be done either by estimating the individual
distances to the stars or by estimating  their scale-height. The first option is quite tricky as
it requires information on metallicity of all stars. The second option is doable with the
existing data and the result obtained can discriminate their disk/halo degeneracy.
Since the LMC
is an almost face-on galaxy, it is a very tricky problem as the stars get projected on the disk
even if they are distributed in a spheroid.
In such cases, indirect methods are used to estimate the scale-height and this is such an attempt
based on the magnitude dispersion of the ab type RRLS. The method was used by
Clementini et al. (2003) to estimate the line of sight depth in their sample.
I have attempted this method with a much larger sample, distributed over a larger area.
The ab type RRLS from the above catalogue is used and their
average magnitude and dispersion as a function of location over the surveyed region,
after correcting for the interstellar extinction was estimated. The depth and the scale-height of their
distribution can be estimated from the observed
dispersion, after correcting for dispersion due to other factors. We also compared
the estimated scale-height with that found for our Galaxy. The results obtained are
discussed in the light of possible formation scenarios.

\begin{figure}
\resizebox{\hsize}{!}{\includegraphics{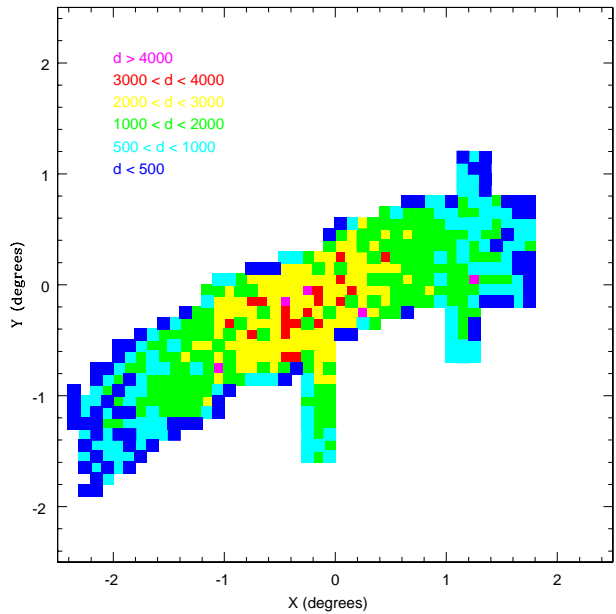}}
\caption{ Number density distribution of RR Lyrae stars in the LMC.
The unit for the number density, d is stars/sq. degree. The colour code is
explained in the figure, where magenta denotes locations of highest density and blue
denotes locations of lowest density. 
\label{fig1}}
\end{figure}

\begin{figure}
\resizebox{\hsize}{!}{\includegraphics{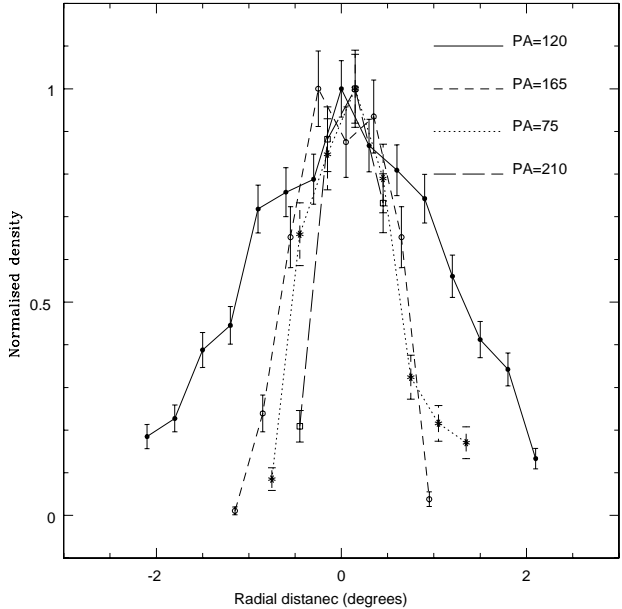}}
\caption{ The normalised number density distribution of RRLS along (filled circles) the 
PA = 120$^o$ (along the bar, filled circle), 30$^o$ ( across the bar, open square),
 75$^o$ (asterix) and  165$^o$ (open circle). The errors include statistical error and a 20\%
more due to any improper incompleteness correction.
\label{fig2}}
\end{figure}

\section{Estimation of number density distribution of RRLS}
\begin{figure}
\resizebox{\hsize}{!}{\includegraphics{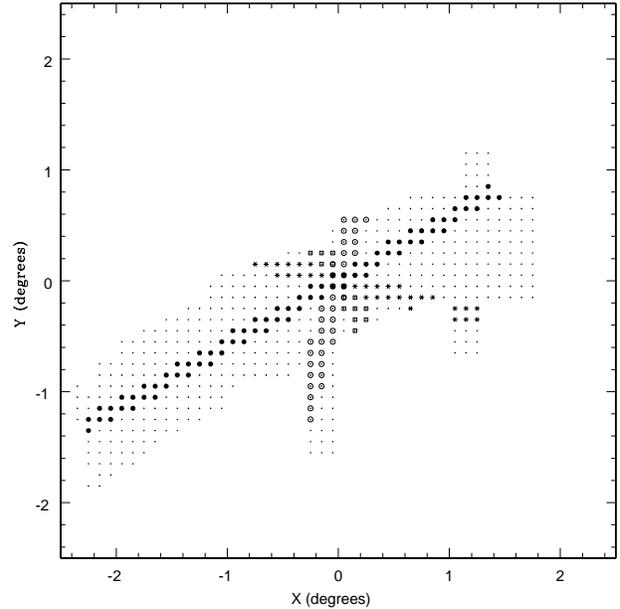}}
\caption{ The locations used to estimate the number density distribution of RRLS as
shown in figure~\ref{fig2}, where 
PA = 120$^o$ (along the bar, filled circle), 75$^o$ (asterix), 165$^o$ (open circle) and
210$^o$ (across the bar, open square).
The small dots indicate spatial coverage of the OGLE II survey.
\label{fig3}}
\end{figure}

Soszynski et al. (2003) presented a catalogue of RRLS discovered in the
4.5 square degree area in the central parts of the LMC, consisting of
7400 objects. The catalogue includes all the sub classes and is more or less
complete ($\sim$ 95 \%). 
The center of the LMC was taken as $\alpha$ = 05$^h$19$^m$38$^s$; 
$\delta$ = $-$69$^o$27'5".2 (de Vaucoulers \& Freeman 1973). The $\alpha$ and $\delta$ were converted to 
the projected X and Y
coordinates. The data are binned in 0.1 degree in both the axes and 
the number of RRLS in each box are counted to obtain the number
density in square degrees. A plot of their density is shown in 
figure~\ref{fig1}, where the variation in the number density is also shown. The 
central regions are found to show higher density, which falls off rather steeply,
approximately along the Y-axis. On the other hand, the density falls
off very slowly towards the sides, which is the X-axis. The density distribution 
shows clumpiness as indicated by the red and magenta
points, some of them denoting the location of old globular clusters, as indicated in
Soszynski et al. (2003). On the whole, the smooth spatial variation in the density
is found to have an elongated distribution. The direction of elongation is found to
be similar to the elongation of the bar of the LMC. Further analysis showed that,
the density distribution along the bar is very different from that across the bar. 
The normalised number density of RRLS along 
four values of PA, (along the bar PA$_{maj}$=120$^o$, across the bar
 PA$_{min}$=210$^o$, and two angles in between,  PA= 75$^o$ and 165$^o$)
 are shown in figure~\ref{fig2}, as a function of radial distance.
The data points used to estimate the above are shown in figure~\ref{fig3}.
These profiles are used to identify and estimate the statistical significance of the
elongated distribution.
If the number density distributions along the two inclined axes are similar, then
this reflects symmetry with respect to the major axis. 
Alves (2004) estimated that the OGLE II
catalogue may be only 80\% complete. This estimation was based on a relative scaling of MACHO
to OGLE II data in the central regions and does not provide any positional dependence.
The error in the density is estimated as the statistical error in the number ($\sqrt N$). 
In order to account for the variation in the density caused by improper completeness estimation,
20\% is added to the above estimated error  and the total error is shown in the figure.

\begin{figure}
\resizebox{\hsize}{!}{\includegraphics{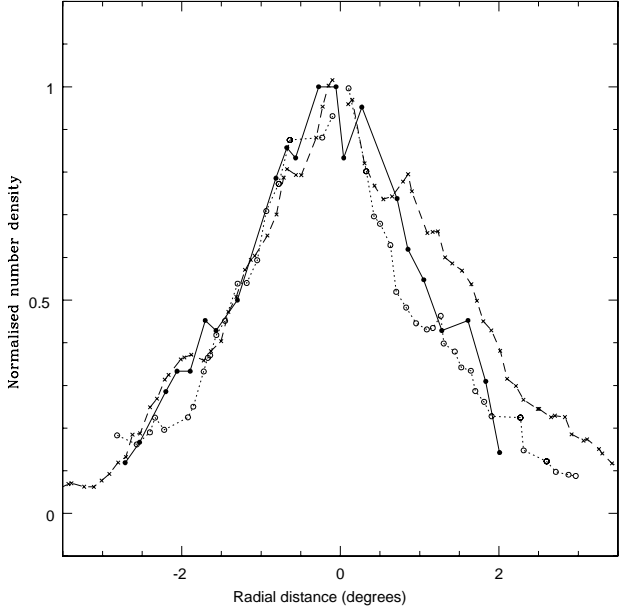}}
\caption{
The normalised maximum number density of RRLS (filled circle),
giant stars (crosses) and RC stars (open circle). The peak value of the distribution are
4300 (RRLS), 1.5 10$^8$ (RC stars) and 2.5 10$^4$ (red giants) stars/sq.degree.
\label{fig4}}
\end{figure}

It can be seen that up to a radius of
$\sim$ 0.$^o$6, there is no difference between the profiles. Thus the distribution of RRLS
is symmetric within this radius. Beyond this radial distance, the profile is found to fall
very slowly along the bar. At a radial distance of $\sim$ 0.$^o$9, the difference in
density along the bar and PA=165$^o$, is more than 5 times the error. This difference continues
up to the radial distance traced here. The profiles along the two inclined 
axes are found to be similar.
The profile along the bar is found to have the maximum width whereas that across the bar has the
minimum width, with the profiles along the inclined axes having intermediate values.
The density curve along the major axis is compared with the profiles along the inclined axes
and a reduced $\chi^2$ value was obtained. The comparison could not be done with
the minor axis as its profile has only 4 points. The reduced $\chi^2$ value for profiles along
the bar and PA = 165$^o$ is 52.1 (8 points), between major axis and PA = 75$^o$ is 44.6 ( 8 points)
 and between the PA = 165$^o$ and 75$^o$ is 9.8 ( 7 points). The large values of reduced $\chi^2$
obtained for profiles along the major axis and the inclined axes indicate that the profile
along the major axis is very different from those along the inclined axes. The elongation of the
density profile along the major axis is thus statistically significant.
The lower value of the reduced $\chi^2$ for the profiles along the inclined axes indicates that
the major axis of the elongation is close to the true major axis. Thus, RRLS
exhibit an elongated distribution in the inner LMC, similar to that exhibited by the bar.

\section{Comparison with RC and giant stars}
The distribution of RRLS is compared with the distribution of
RC stars and red giants, and the bar signature as revealed by these populations are compared.
The density distribution of the RCs was studied
by Subramaniam (2004) and found that it showed not only the bar feature of the LMC, but also indicated
the presence of a secondary bar. In general, the RC stars can have a wide range of ages, between
2 -- 10 Gyr. In the LMC, 
the bulk of the RCs are estimated to be a few Gyr old Girardi \& Salaris (2001), and they
represent the intermediate age population.
The density distribution of the giant stars located in the upper part of the 
red giant branch and in the asymptotic giant
branch was studied by van der Marel (2001) and he also estimated the parameters of the bar such as,
PA and ellipticity. This stellar population has a broader age range than
the age range of the RCs in the LMC.
The data presented in their figure 2c was used for the
present analysis. For all the three data sets, the maximum density points as a function of
radial distance was estimated from the east to the west. The method used is described in
Subramaniam (2004). The results for the RCs were already
presented in their figure 1. A similar plot for the RRLS is shown here in
figure~\ref{fig4}. While estimating the above, care was taken to avoid points
corresponding to the locations of GCs.
The profile is found to have a peak and broad wings, with the peak shifted from the center.
The distribution of RRLS is found to have a preferred PA and this indicates a
(projected) non-circular distribution. The average value of the PA 
is estimated as PA = 112.5$^o$ $\pm$ 15.$^o$3.
The above value is very similar to the PA$_{maj}$ as estimated from red clump stars, 114$^o$.4 $\pm$ 22$^o$.5,
(Subramaniam 2004) and also, the PA$_{maj}$ from the red giants 122$^o$.5 $\pm$ 8$^o$.3 (van der Marel 2001). 
All the above three values are the same within the errors.
The density distributions of all the three populations are normalised for comparison 
and are shown in figure~\ref{fig4}.
In the east, all the three distributions match very well. In the west, the three distributions
differ slightly.
Thus, the density distribution of RRLS is very similar to that of younger population along the 
major axis.
This result indicates that the scale length of the RRLS along the major axis
is similar to the scale length of the younger population. 
Alves (2004) estimated the scale length of the RRLS as 1.47 $\pm$ 0.08 kpc, which
is very similar to the scale length of LMC's blue light, 1.46 kpc (Bothum \& Thompson 1998).
We also find that the scale length of the RRLS is very similar to that of RC and red giant stars,
which is about $\sim$ 1.3 kpc (1 degree = 0.89 kpc) as obtained from figure 4.
The bar feature as shown by the RRLS can be compared to that shown by the
younger population, also by using ellipticity. The ellipticity is estimated by
van der Marel (2001), as $\epsilon$ = (1$-$b/a). 
The value of semi-major and semi-minor axes are estimated from figure ~\ref{fig2}. 
The value of $\epsilon$ was found to be very close to zero, up to a radial distance of $\sim$ 0.$^o$6. 
Between the radial distance of 1.$^o$ -- 1.$^o$8, the value of $\epsilon$ was found to increase
from 0.45 to 0.57, with an average value of $\epsilon$ $\sim$ 0.5.
van der Marel (2001) estimated that the red giants in the inner r=3$^o$, showed ellipticity value
in the range 0.43 to 0.67.  
Thus the distribution of both RRLS and red giants show
similar values for ellipticity. In summary, the bar type elongation shown by the RRLS
is similar to that delineated by RC stars and red giants. 
The density distribution of the RRLS is found to be located slightly away
from the adopted optical center of the LMC. Their center was estimated by
centering the density distribution along the major axis and was found to be
$\alpha$ = 05$^h$ 20$^m$.4$\pm$0.$^m$4 and $\delta$ = $-$69$^o$ 48'$\pm$5'. This is the same
as estimated by Alves (2004) within the errors.

The result obtained above indicates that the distribution of RRLS, has an imprint
of the bar feature. A simple interpretation of the above result
is that these are located in the disk (assuming the bar to be a disk feature), 
similar to the red clump stars and
red giants. On the other hand, RRLS are expected to belong to an older
population ($\ge$ 9 Gyr) and hence the majority are expected to be located in the halo, though their location
in the LMC is still uncertain.  Since the distribution estimated here is the projected one,
it does not straight away establish their location in the disk. For a disk location, 
one expects a  very small depth in the line of sight
(scale-height) in their distribution and thus their scale-height can be used to verify the above claim.
In the following section, we estimate the average scale-height of their
distribution, where the value of the scale-height is used as a pointer towards deciding 
the location of RRLS.
                                                                                                       
\section{Estimation of dispersion in the mean magnitude of ab type RRLS}
The catalogue used here has 5077 ab type RRLS. These stars could be considered
to belong to similar sub class and hence assumed to have similar properties.
The mean magnitude of these stars in I passband, after correcting for the metallicity and extinction
effects can be used for the estimation of distance. On the other hand, the observed
dispersion in their mean magnitude is a measure of the depth in their distribution.
Subramaniam (2005) presented a high-resolution reddening map of the region
observed by the OGLE survey. This data is used to estimate the extinction to individual
RRLS. Stars within each bin of the reddening map are assigned a single reddening
and it is assumed that the reddening does not vary much within the bin. The contribution
to the dispersion from the variable extinction is minimised, though there may still be a
non-zero contribution to the estimated dispersion.

Since the extinction corrected mean magnitudes of 5077 ab type RRLS are available
in the surveyed region, it is possible to study the variation of the dispersion as a
function of location in the inner LMC. This data can be used to create a map of depth
across the LMC, where the corrected magnitudes are assumed to be a measure of the distance.
This may not be strictly correct, since the relation depends on the metallicity and age which is
not accounted for. The picture, on the other hand, will be able to give a rough idea of the
location of these stars, without giving any quantitative result. 
The extinction corrected magnitudes are plotted as a function of RA as shown in
figure 5.
\begin{figure}
\resizebox{\hsize}{!}{\includegraphics{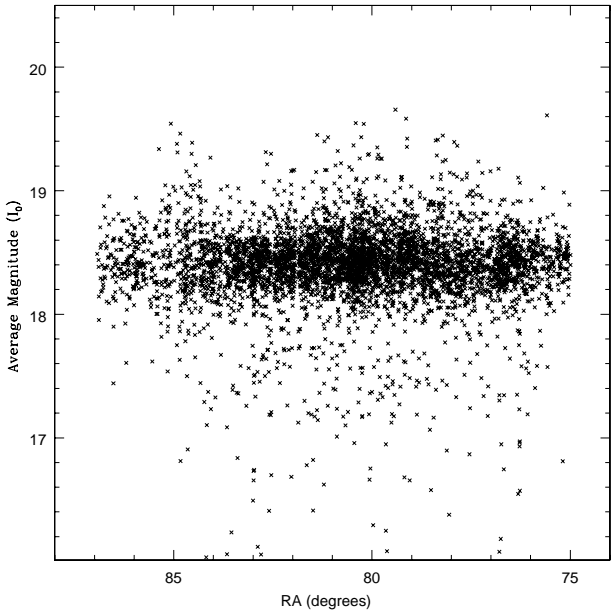}}
\caption{The variation in the extinction corrected magnitude as a function of RA.
Note that the distribution is symmetric with respect to a concentration near $I_0$ = 18.45 mag, which
can be considered as the disk of the LMC.}
\label{fig5}
\end{figure}
It can be seen that there is a range in the location of RRLS with respect to
the disk of the LMC. 
Note that the distribution is symmetric with respect to a concentration near $I_0$ = 18.4 mag, which
can be considered as the disk of the LMC.
Though many stars are located closer to the disk, there is a significant
number located away as well. Also, the plot shows that RRLS located to the front
and back of the disk are detected. The plot definitely indicates that RRLS are located
closer as well as far away from the disk indicating an elongated distribution. There are
more RRLS in front of the disk and closer to us then those behind the disk. Does this indicate
that the RRLS distribution is stretched towards our Galaxy? We shall discuss this in more detail
in the following sections. To obtain a quantitative estimate of the scale-height, after correcting for the
biases present in the above plot, we follow the procedure described below.
The observed distribution of ab type stars
in RA and Dec are binned such that there are more than 10 stars in most of the location and with
a maximum of $\ge$ 60 stars in a some locations. We used a bin size of 0.$^o$5 in RA
and 0.$^o$2 in Dec. We estimated the mean magnitude and dispersion for 163 locations.
This value of the dispersion can be used to estimate the scale-height, but
has contributions from (1) photometric errors, (2) range in
the metallicity of stars, (3) intrinsic variation in the luminosity due to evolutionary
effects within the sample and (4) the actual depth
in the distribution of the stars (cf Clementini et al. 2003). We need to remove the
contribution from the first three terms ($\sigma_{(int)}$) so that the value of the
last term ($\sigma_{(dep)}$) can be evaluated.
Clementini et al. (2003) estimated the value of $\sigma_{(int)}$ as 0.1 mag for their sample.
The second and third terms could be considered similar for any set of data, since it is intrinsic
to RRLS, whereas the first term depends on the data set. Therefore, we need to get
an independent estimate of $\sigma_{(int)}$ for the data sample used here. The
catalogue used here also has RRLS detected in a few clusters in the bar 
region of the LMC. The RRLS
located in a cluster can be assumed to be located at the same distance and will not
have the contribution from the depth in the distribution. Also, the contribution from the
second term in the above list can be considered to be minimal. If we estimate the average
and the dispersion of the mean magnitudes of the ab type RRLS in a cluster, the
estimated value of the dispersion will have contribution mainly from the first and third
terms. The value of the dispersion estimated
for NGC 1835 (based on 27 stars) is 0.14 mag, NGC 2019 (30 stars) is 0.15 mag and
NGC 1898 (31 stars) is 0.17 mag. Clementini et al. (2003) estimated that the dispersion
due to metallicity could be 0.06 mag. If we assume this value to be typical, then this term
can be combined with the estimates from the cluster to give an estimate of $\sigma_{(int)}$.
Thus we find the lower limit of the value to be 0.15 mag and the
upper limit to be 0.18 mag. Thus the value of $\sigma_{(int)}$ for the present data set is slightly higher
than that estimated by Clementini et al. (2003) for their data. The fact that the values are
not very different indicates that the present estimate of $\sigma_{(int)}$ is
likely to be closer to the actual value.

\begin{figure}
\resizebox{\hsize}{!}{\includegraphics{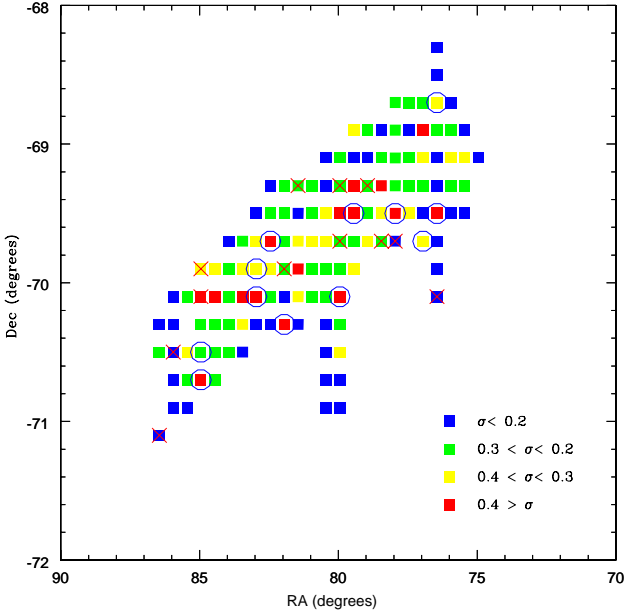}}
\caption{2-D plot of the estimated value of dispersion ($\sigma_{(dep)}$) in the mean magnitude of the
ab type RRLS. The colour code is indicated in the figure. The open circles indicate locations which have
average mean magnitude less than 18.3 and crosses denote location which have values greater than 18.5 (see 
figure 7). }
\label{fig6}
\end{figure}
It turns out that the observed dispersion in many regions is
less than the upper limit of the error as indicated above. Therefore, we used
the value for the lower limit (0.15 mag) in our analysis. Also, as the contributions are added quadratically,
the variation introduced in the estimated value of dispersion due to the above assumption
is very small. The value of $\sigma_{(dep)}$ thus obtained directly indicates the depth in the distribution.
A 2-D plot of the observed dispersion is shown in Figure 6.
It can be seen from the figure that there is a significant variation in the value of
$\sigma_{(dep)}$ across the observed region. The difference between
the red and the blue points is greater than 0.2 mag. For this difference to be just due to an enhanced
value of $\sigma_{(int)}$, it has to be  more than 0.35 mag. This is more than twice the
value of the maximum dispersion as estimated from the cluster RRLS. The increased dispersion
is  unlikely to be due to the variation in the extinction, since the
reddening (and thus extinction) was not found to show significant variation across the bar (Subramaniam 2005).
                                                                                                                    
\begin{figure}
\resizebox{\hsize}{!}{\includegraphics{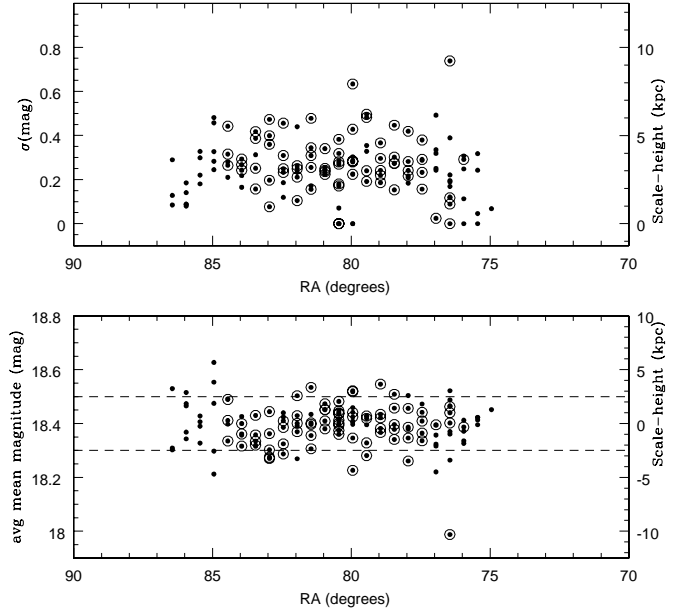}}
\caption{The variation of $\sigma_{(dep)}$ (upper panel) and scale-height (lower panel)
 as a function of RA. Locations with number of stars more than 10 are
shown as dots and those with more than 30 stars are shown with circles around them. }
\label{fig7}
\end{figure}
\begin{figure}
\resizebox{\hsize}{!}{\includegraphics{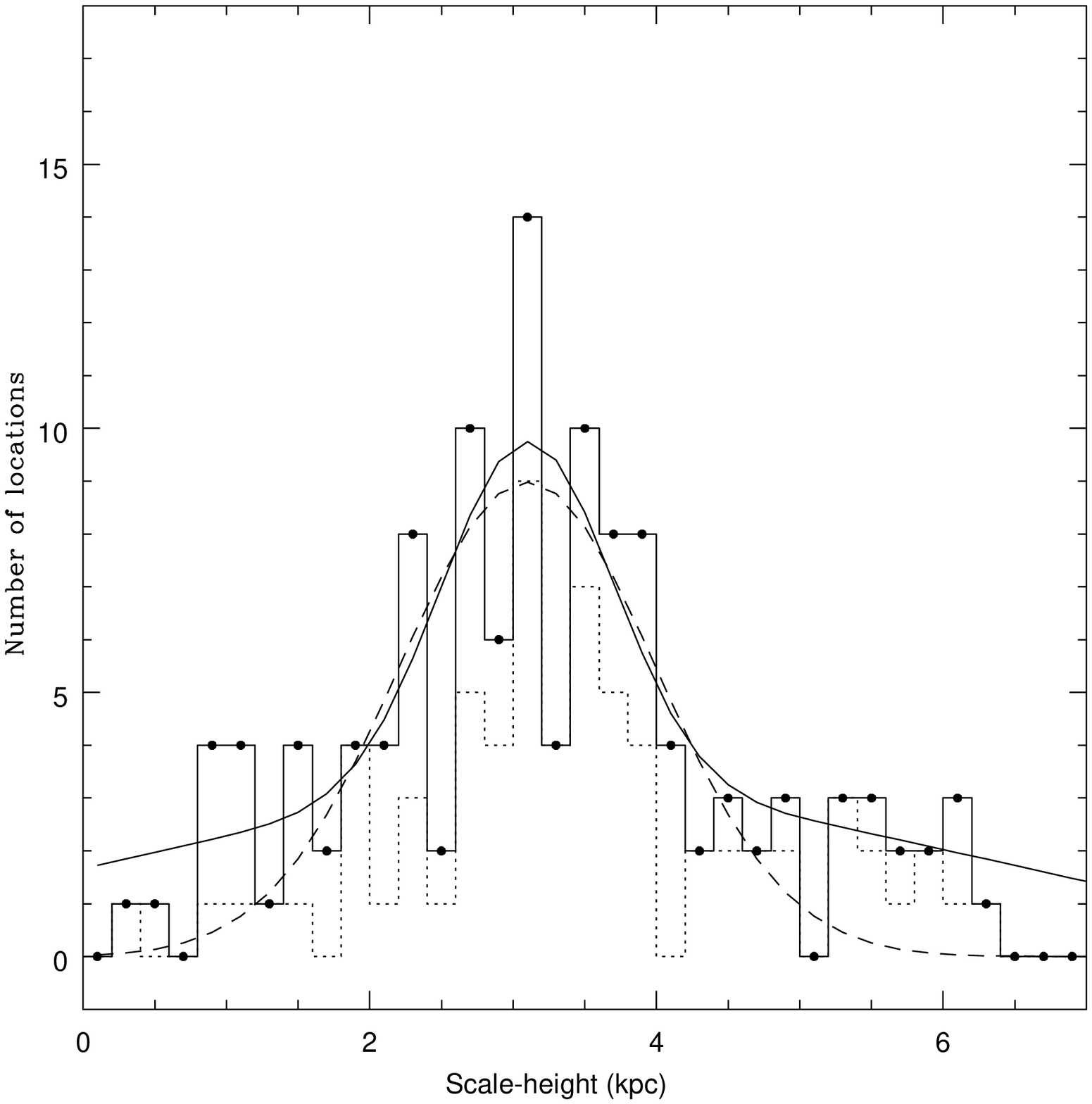}}
\caption{The distribution of the scale-height. The solid line is for
locations with more than 10 stars and dotted line is for locations with more than 30 stars.
Single Gaussian fit is indicated by dashed curve and double Gaussian fit is indicated by solid curve. }
\label{fig8}
\end{figure}

Since the surveyed region has more coverage along the RA,
the dispersion thus obtained can be plotted
a function of RA and can be used to understand the variation in scale-height across the bar.
This will also correspond to the edge on view of the bar of the LMC. The value of dispersion
is found to take a range. The average value of $\sigma_{(dep)}$
is found to be 0.24 mag, if we assume a value of 0.15 mag for $\sigma_{(int)}$.
In this case we considered 133 locations with more than 10 stars.
The value becomes 0.22 mag, if we assume a value of 0.17 mag for $\sigma_{(int)}$. It can also
be seen that the upper limit for $\sigma_{(dep)}$ is 0.45 mag, as shown in the upper panel of figure 7.
A Gaussian fit to the distribution was used to estimate the peak and the width of $\sigma_{(dep)}$.
Thus the average value of $\sigma$ is 0.25$\pm 0.07$ mag. If the whole sample of RRLS is considered,
then the estimated value of $\sigma$ is found to be 0.28 mag.
In order to show that this is not a statistical effect, we have shown locations with more
than 30 stars separately (encircled dots). These locations also show significant range in $\sigma_{(dep)}$,
proving that it is not due to poor number statistics within a given location. The average
of the observed $\sigma_{(dep)}$ for locations with more than 30 stars is 0.29 mag (76 locations).
Thus all the above values are found to be within the errors.
In the lower panel, we have shown the variation of the mean I magnitude of the ab type RRLS
as a function of RA. Since the mean magnitude indicates the distance to the LMC, the plot
shows that there is no significant variation in distance along the bar.  Most of the locations have
magnitudes between 18.3 and 18.5 mag where the mean was found to be 18.40 mag. 12 locations were
found to be located to the front and behind the above range as their magnitudes were found to be
brighter and fainter respectively (see figure 6). 
On the other hand, a significant amount of scatter can be found even for
a particular value of RA. One could argue that the
variation seen in the the mean magnitude as well as the dispersion is due to
significant differential extinction, more than what has been corrected for.
This is unlikely if the RRLS are halo stars
and high differential reddening is normally expected for young disk population.
The reddening used here was derived from RC stars which belong to the disk population.
Therefore, the reddening estimated using RC stars can still be valid for RRLS even if they
belonged to the disk population. Though the estimated elongation is genuine, there may be
a non-zero, but small contribution due to differential reddening.
The other possibility is that these stars are not distributed uniformly,
but in clumps. This also can give rise to a range in mean magnitudes and dispersion.

\section {Estimation of scale-height}
The value of $\sigma_{(dep)}$ estimated above can be converted to the scale-height.
If we assume that the depth in the distribution in this almost face-on galaxy is symmetric
with respect to the disk (that is, one half coming from behind the disk and the other half
from the front), then $\sigma_{(dep)}$/2.0 will correspond to the scale-height in terms
of magnitude. The value of $\sigma_{(dep)}$ can be converted to distance assuming the relation,
0.1 mag $\sim$ 2.5 kpc, for a distance modulus of 18.50 mag. Thus we define
scale-height as $ (\sigma_{(dep)}/2.0) \times 25.0$ kpc. The scale-height is shown
in the right side of the y-axis in figure 7 (upper panel).
In order to get a feel for the number distribution of the scale-height, we have
plotted the histogram of the estimated scale-height in figure 8.
The solid line shows the distribution for locations with more
than 10 stars and dotted line shows the distribution for locations with more than 30 stars.
It can be seen that the peak of the distribution is at 3.0 kpc for both the plots and most
of the locations have scale-height in the range of 2.0 -- 4.0 kpc.
A single Gaussian fit to the
distribution, as shown by dashed lines estimated a value of 3.1$\pm$0.9 kpc as the 
scale-height. It can be seen that the fit is poor (reduced $\chi^2$ = 5.7) due to the
presence of deviating points at the wings of the Gaussian. A double Gaussian
fit was attempted and the fit thus obtained (reduced $\chi^2$ = 4.1) is shown in the figure.
The values of the peak, which is the scale-height was found to be similar (3.1 and 3.3 kpc),
whereas the error in the scale height was found to be 0.6 and 3.0 kpc respectively.
The estimated scale height has two components, most of the locations show a smaller spread,
whereas a small number of locations show a larger spread. Let us see whether there is any preferred
location for the points with higher scale-height. Of the 12 brighter locations,
as shown in figure 6, 11 locations (92\%) were found to have scale-height more than 3.75 kpc
and 8 locations (66\%) were found to have scale-height larger than 5 kpc. On the other hand,
of the 12 fainter locations, 10 (83\%) were found to have scale-height less than 3.75 kpc. Thus
it appears that the locations in front of the disk have larger scale height when compared to those
behind the disk.
The above  numbers (fractions) quantifies the indication of elongation of the RRLS system towards our
Galaxy. The stretching of the halo of the LMC towards the Galaxy is expected due to tidal effects.
Thus this may be the first evidence of such an effect on the LMC halo.

Clementini et al. (2003) estimated a depth of 3.3 -- 3.8 kpc which is same as a scale height
of 1.6 -- 1.9 kpc. It can be seen from figure 1 and 2 that the value of the scale height can
vary depending on the location. Thus the present
estimate is not in disagreement with the results of Clementini et al. (2003).
In order to estimate the scale-height of the distribution, we assumed that the observed
dispersion has equal contribution from the front and back of the LMC disk.
This may not be strictly true. Since the disk is face on, there may be incomplete
detection of RRLS located behind the disk.
Thus instead of half of $\sigma_{(dep)}$ contributing to the scale-height, a higher fraction
should be contributing to the scale-height. It is not possible to estimate the value of
the fraction to be used, as it requires an estimate of incompleteness in the detection of RRLS
located in the back, which is not available. The net result being, the
above assumption leads to a lower value of the scale-height. Thus the value of scale-height
estimated can be considered to be a lower limit. If we assume that all stars are located in front
of the disk, then the scale-height is twice the value given above, which will be 6.0 $\pm$ 1.2 kpc.
In reality it is likely to be less than this, indicating these
values are the upper limits.

The majority of RRLS indicate that the scale-height of their distribution is 3.1$\pm$0.9 kpc.
This indicates that they are not located on the LMC disk,
but in a halo, even though their distribution mimics that of the LMC disk.
The value of their scale-length as estimated from figure 4 is found to be $\sim$ 1.3 kpc.
This value is estimated from the number density distribution. If we assume that the scale-length and the
scale-height estimated here can be directly compared, then
RRLS are elongated more along the line of sight than along the major axis of the disk.
The characteristics of this halo needs to be studied in detail to understand the early
formation history of the LMC.

\section{A flattened LMC halo?}
We compared the value of scale-height with that estimated for our Galaxy.
Layden (1995) estimated a value of 2.7$^{+10}_{-2.0}$ kpc for
the scale-height of RRLS
in our Galaxy. Thus we find that the average value of the scale-height estimated for the LMC
closely resembles the value obtained for our Galaxy.
Using the estimated scale-height, Layden (1995) calculated the probable space
density of RRLS. He found that it is consistent with a flattened halo (c/a $\sim$ 0.5).
Thus it is likely that the LMC, with a similar value for the scale-height, could also have a
flattened halo. Kinman et al. (1996) estimated the velocity dispersion
of 9 RRLS with Z $\le$ 4 kpc in our Galaxy as 40 $\pm$ 9 kms$^{-1}$ and
remarked that this is
expected if the population belongs to a flatter distribution. The observed velocity dispersion
of 53 $\pm$ 10 kms$^{-1}$ for RRLS in the inner LMC by Minniti et al. (2004) is
similar to the value indicated above. Gratton et al. (2004) observed a velocity dispersion of
30 kms$^{-1}$ among 48 stars. Thus, the above values of
dispersion are consistent with the arguement that the inner LMC has a flattened halo.
Thus it is very likely that the inner LMC has a halo very similar to that of our Galaxy,
at least within Z $\le$ 5 kpc.
This is a very significant result with respect to the galaxy formation scenarios.
Galaxy formation models by dissipational collapse ( Eggen, Lynden-Bell \& Sandage 1962)
predict the presence of old stellar halo for all galaxies,
irrespective of their size. The result presented here indicates that the LMC has at least a halo
which is similar to the inner halo of our Galaxy. The outer halo of LMC,
which is characterised by a spatially extended old population with
large velocity dispersion ($\sim$ 100 kms$^{-1}$), is yet to be detected.

\section{Result and Discussion}
The analysis presented here showed that the number density distribution of the
RRLS differs along and across the bar, indicating a preferred elongation
along the major axis of the bar. The bar feature as shown
by RRLS is very similar to that shown by the younger population, in terms of
PA$_{maj}$, ellipticity and variation of density along the major axis. This indicates that
the bar feature as seen in the disk is also found in the RRLS. One of the possible
interpretation of this result is that these stars are located in the disk and 
not in the halo and hence they mimic the bar feature. This interpretation is
against the general belief that the majority of RRLS belongs to the halo population.
We estimated the scale-height of the ab type RRLS and was found to be
 3.1$\pm $0.9 kpc, indicating that the majority are not located in the disk, 
but in the halo, which may be flattened. Therefore, 
the RRLS in the LMC are in fact located in the halo. Thus RRLS in the inner LMC have
a strange combination of a disk-like
density distribution and halo-like location. We also find marginal evidence for stretching of the LMC halo
towards our Galaxy.

The scenario described by RRLS in the inner LMC seems to show the signatures
of both disk as well as halo. This result is quite puzzling. This is similar to a situation
where a spheroidal system, mimics the feature of a flattened disk-like system. At this stage
it is not very clear as to what could be the origin of such a system. Detailed study of 
RRLS including their kinematics and metallicity is required to solve this puzzle.

Here we discuss some possibilities for obtaining such systems, in general.
If RRLS were formed in the halo, then they will most probably have a symmetric
density distribution with a non-disk like luminosity function. After it is formed, it is not
possible to create the above imprints of the disk on an already existing spheroidal
distribution. Thus this scenario is unlikely. 
If we assume that RRLS were initially formed in the disk, and then
the distribution is quite likely to have the disk properties as obtained here. 
Such a distribution can get dislocated into a puffed up distribution.
This would relocate the stars into the halo (away from the disk) but still maintaining the
disk properties in terms of density and luminosity. Thus this scenario can explain the
observed properties. This kind of situation can arise in mergers, where the disk gets 
heated up and results in a puffed up distribution. Jog \& Chitre (2002) found some galaxies
which have exponential light profile (like the disk) and kinematics indicating pressure
support rather than rotation (puffed up location). Bournaud et al. (2004) found that 
mergers of intermediate mass galaxies (4:1 - 10:1) results in similar such peculiar systems.
These type of mergers are more common  at high red shift as shown in the
hierarchical merging models (Steinmetz \& Navarro 2002) for galaxy formation. Combining
all the above information, it is likely that the LMC went through mergers in the 
mass range as indicated above in its early formation. Recent interactions of the LMC
with our Galaxy can induce tidal effects on the LMC, which has been already noticed in the disk by
van der Marel (2001). The halo of the LMC can also get stretched towards our Galaxy due to tidal effects.
Thus interaction and merger events may have resulted in the present distribution of RRLS.
\begin{acknowledgements}
I thank the referee Dr. Girardi for constructive suggestions and
Abhijit Saha, Chanda J. Jog and Tushar P. Prabhu for fruitful discussions.
\end{acknowledgements}


\begin{thebibliography}{}
\bibitem[Alcock et al.(1996)]{alk96} Alcock et al. 1996, \aj, 111, 1146
\bibitem[Alves(2004)]{al04} Alves, D.R. 2004, \apjl, 601, L51
\bibitem[Bothum \& Thompson (1998)]{bt98} Bothum, G.D., \& Thompson, I.B., 1988, AJ, 96, 877
\bibitem[Bournaud et al.(2004)]{b04}Bournaud, F., Combes, F., Jog, C.J., 2004, A\&A, 418, L27 
\bibitem[Clementini et al.(2003)]{C03}Clementini et al. 2003, AJ, 125, 1309
\bibitem[de Vaucoulers \& Freeman(1973)]{df73} de Vaucouleurs, G., \&
  Freeman, K.C. 1973, Vistas Astron., 14, 163
\bibitem[Eggen, Lynden-Bell \& Sandage(1962)]{els62}Eggen, O., Lynden-Bell, D., \& Sandage, A., 1962,
  \apj, 136, 748 
\bibitem[Freeman(1999)]{f99} Freeman, K.C., 1999, IAU Symposium 192, Patricia
 Whitelock and Russell Cannon eds., p383 
\bibitem[Girardi \& Salaris(2001)]{gs01} Girardi, L., \& Salaris, M., 2001, MNRAS, 323, 109
\bibitem[Gratton et al.(2004)]{gr04} Gratton, R.G., et al. 2004, A\&A, 937, 952
\bibitem[Jog \& Chitre(20020]{jc02} Jog,C.J. \& Chitre, A., 2002, A\&A, 393, L89
\bibitem[Kinman et al.(1991)]{k91} Kinman, T.D., Stryker, L.L., Hesser.J.E.,
 Graham, J.A., Walker, A.R., Hazen,M.L., \& Nemec, J.M., 1991, PASP, 103, 1279
\bibitem[Layden(1995)]{lay95} Layden, A.C., 1995, \aj, 110, 2288
\bibitem[Minniti et al.(2003)]{m03} Minniti, D., Borissova, J., Rejkuba, M., Alves D.R.,
Cook, K.H., \& Freemen, K.C. 2003, Science, 301, 1508
\bibitem[Schommer et al.(1992)]{sch92} Schommer, R.A., Olsewski, E.W., Suntzeff, N.B.,
\& Harris, H.C. 1992, \aj, 103, 447
\bibitem[Soszynski et al.(2003)]{sz03} Soszynski, I., et al. 2003, Acta Astron., 53, 93
\bibitem[Steinmetz \& Navarro(2002)]{sn02}Steinmetz, M. \& Navarro.J.F., 2002, New Astron., 7, 155
\bibitem[Subramaniam(2004)]{as04} Subramaniam, A. 2004, \apjl, 604, 41
\bibitem[Subramaniam(2004)]{as05} Subramaniam, A. 2005, A\&A, 430, 421
\bibitem[van den Bergh(2004)]{vb04} van den Bergh, S., 2004, \aj, 127, 897
\bibitem[van der Marel(2001)]{v01} van der Marel, R.P. 2001, \aj, 122, 1827
\end{thebibliography}
\end{document}